\documentclass[preprintnumbers,amsmath,amssymb]{revtex4}

\usepackage{graphicx} % Include figure files
\usepackage{times}
\usepackage{amsmath}
\usepackage{amssymb}
\usepackage{units}

\usepackage{dcolumn}% Align table columns on decimal point
\usepackage{bm}% bold math

\begin{document}
\title{Interferometer-controlled soft X-ray scanning photoemission microscope at SOLEIL}

\author{Jos\'{e} Avila, Ivy Razado-Colambo, Stephane Lorcy, Jean-Luc Giorgetta, Fran\c{c}ois Polack and Mar\'{i}a C. Asensio}

\address{Synchrotron SOLEIL\\ Orme des Merisiers - Saint Aubin\\ 
BP 48 - 91192 - GIF SUR YVETTE Cedex\\  FRANCE}

\begin{abstract}

ANTARES beamline (BL), a new soft X-ray scanning photoemission microscope located at the SOLEIL synchrotron storage ring has been recently designed, built and commissioned. The implemented interferometer control allows the accurate measurement of the transverse position of the Fresnel zone plate (FZP) relative to the sample. An effective sample position feedback has been achieved during experiments in static mode, with a fixed FZP position required to perform nano Angle-Resolved Photoelectron Spectroscopy (Nano-ARPES) measurements. Likewise, long-term stability has been attained for the FZP position relative to the sample during the translation of the FZP when performing typical X-ray absorption experiments around the absorption edges of light elements. Moreover, a fully automatic feedback digital control of the interferometric system provides extremely low orthogonal distortion of the recorded two-dimensional images. The microscope is diffraction limited with the resolution set to several tens of nanometers by the quality of the zone plates. Details on the design of the interferometric system and a brief description of the first commissioning results are presented here.

 \end{abstract}

\maketitle

\section{Introduction}

In a scanning X-ray microscope, a small spot of X-ray is raster scanned relative to the sample to create an image one pixel at a time while a suitable signal is monitored under computer control. In ANTARES, the signal is recorded by a high energy resolution Scienta R4000 spectrometer that ensures a good energy- and angle-resolved photoelectron detection in the range of 10 eV to 1000 eV. A specially designed BL optics guarantees a homogeneous and coherent illumination of a set of FZPs, which focalizes the beam spot up to a few tens of nanometers, depending on the performance of the FZP being utilized. The BL is completed with an advanced Nano-ARPES microscope [1], with precise sample handling capabilities, non-magnetic and fully compatible with ultra high vacuum conditions. The piezo stages allow not only a precise mechanical scanning and positioning of the samples but also a fast and highly reproducible instrument alignment with respect to the photon beam and analyzer focus. The microscope consists of fourteen independent axis manipulators which focus the beam and perform sample scanning with nanometric accuracy. The sample position relative to the focal distance of the FZP is controlled with high precision during experiments at a constant or variable photon energy by a continuous interferometric monitoring and an automatic feedback mechanism  [2].  

\begin{figure}[h]
\includegraphics[width=27pc]{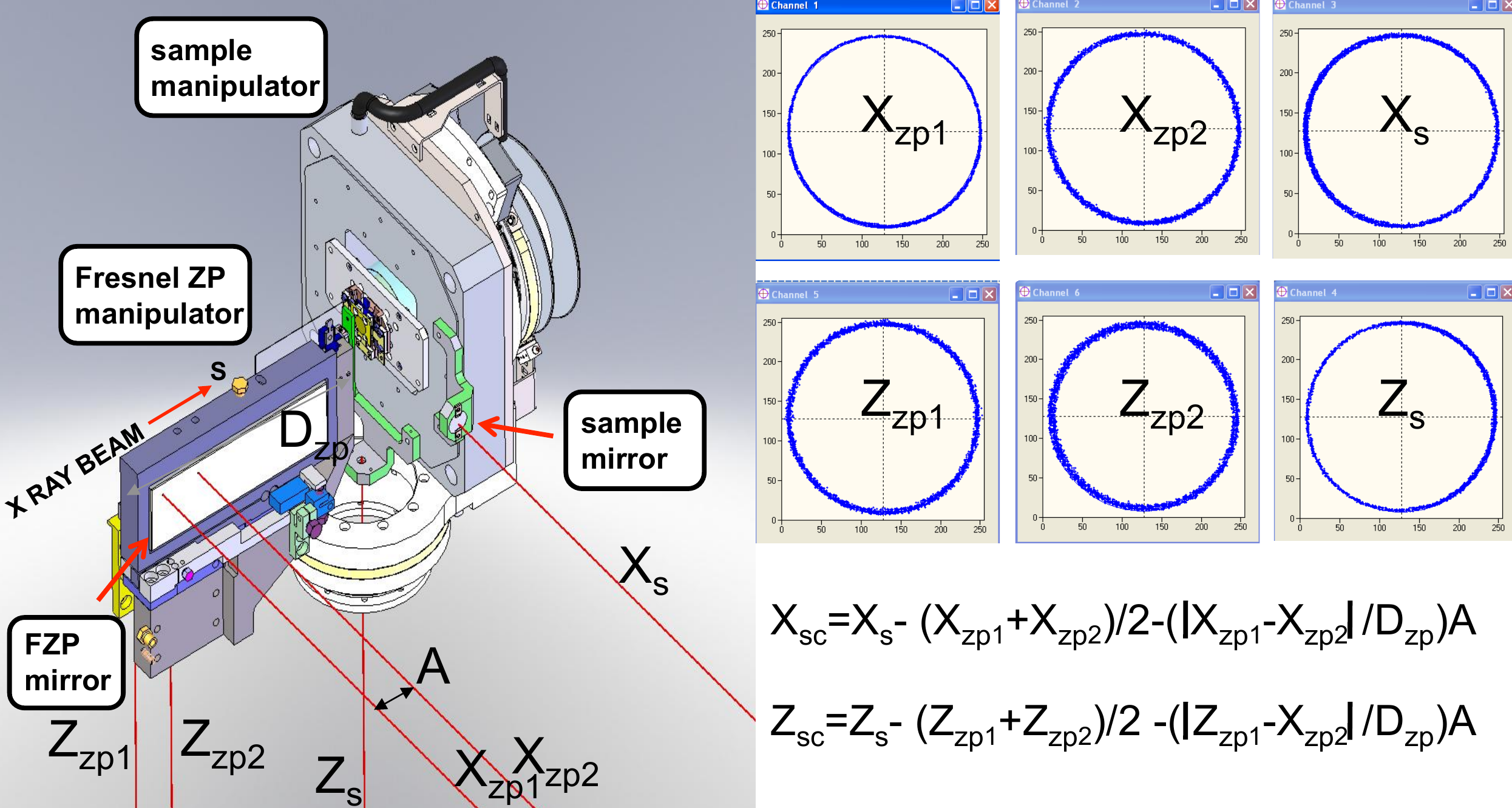}\hspace{2pc}%
\begin{minipage}[b]{27pc}\caption{\label{label} The left panel is a CAD drawing  showing the monolithic manipulator that holds the sample and the FZP  mounted in the ANTARES microscope. Six laser beams along the horizontal (x-s) and vertical (z-s) planes (s is the direction of the incident X-ray beam) are reflected on the sample and FZP mirrors. The top right panel displays the standard quality of the six individual interferometer signals as Lissajous curves corresponding to the X and Z positions of the sample and the FZP, respectively. The equations at the bottom right of the figure indicate the corrected X$_{sc} $ and Z$_{sc} $  positions of the sample which are calculated as a function of the interferometric signals. }
\end{minipage}
\end{figure}

\section{Interferometer }

Direct interferometric differential measurement of the position of the FZP with respect to the sample is implemented in the ANTARES microscope, (see Fig. 1). Three laser beams are placed in the horizontal (x-s) plane. Two of them are dedicated to measure the X coordinate of the FZP and its angular deviation. The third horizontal laser beam records  the X coordinate of the sample. Three additional laser beams are placed in the vertical plane to get the equivalent information in a plane (z-s) perpendicular to the horizontal one.  Two double and two single beams of the SP-series SIOS miniature plane-mirror interferometers (Models SP 120/500/2000) are used with the mirrors mounted to the FZP carrier and the sample, respectively. The precision of this configuration is 2.5 nm. These laser-interferometers are permanently placed in a dedicated vacuum chamber close and independent to the analysis chamber. The beam from the He-Ne laser source is transmitted to its respective sensor head by a fiber optic cable.  The interferometers convert motions of their planar mirror along the beam axis into optical interference signals that are transmitted to their optoelectronic signal processing/power supply units for processing and output as lengths. The use of a closed-loop piezo stage motion with a differential interferometer as a feedback is one of the major reasons for the remarkable stability and reproducibility of the ANTARES microscope. 

\begin{figure}[h]
\includegraphics[width=27pc]{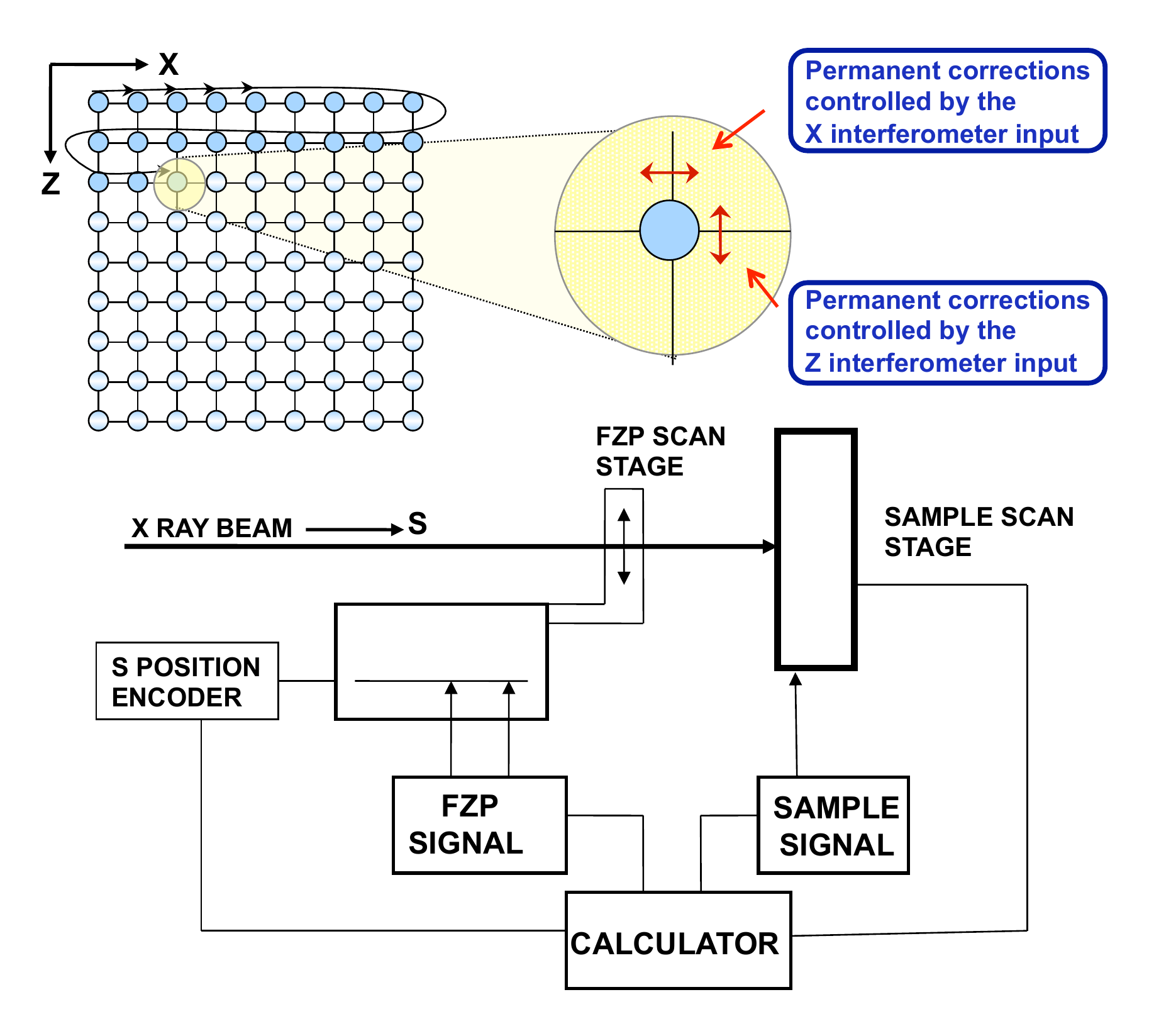}\hspace{2pc}%
\begin{minipage}[b]{27pc}\caption{\label{label}  Schematic diagram of the step by step image generation and the control system of the ANTARES microscope with a differential interferometric feedback closed-loop.}
\end{minipage}
\end{figure}

\section{Electronic interface}

Figure 2 is a simplified schematic diagram of the fully automated  ANTARES microscope control system. The differential interferometers measure the horizontal and vertical positions (x and z) of the sample with respect to the FZP lens by means of mirrors attached to these components which are illuminated by the laser beams. Hence, the interferometers provide an accurate real-time measurement of the location of the X-ray spot on the sample. Digital signals from the interferometer are used to servo the high precision sample piezo-scanning stage, in order to eliminate drift and mechanical vibrations under both operating modes; the static mode with the FZP fixed during the experiment, and the scanned mode where the FZP is scanned during the image acquisition.

\begin{figure}[h]
\includegraphics[width=27pc]{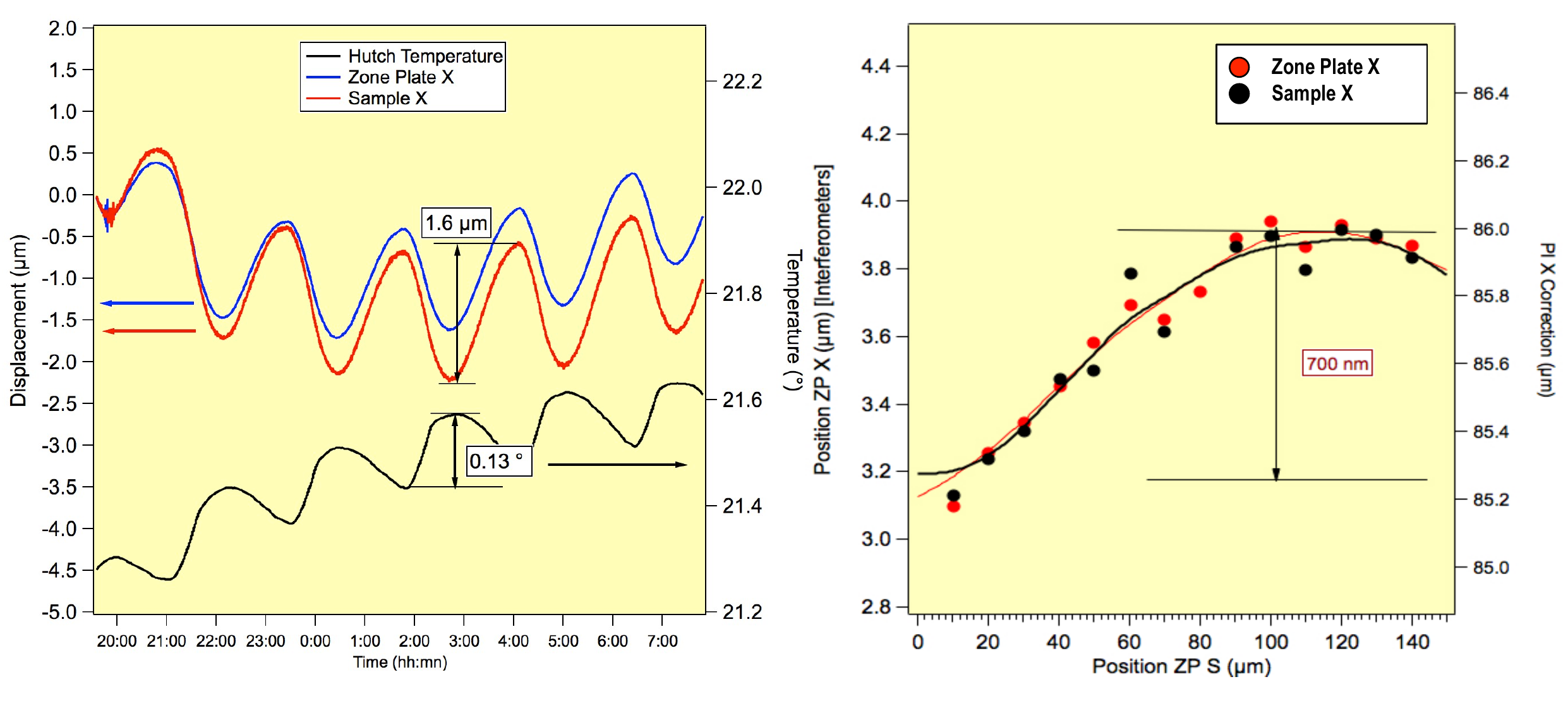}\hspace{2pc}%
\begin{minipage}[b]{27pc}\caption{\label{label} Mechanical vibrations and thermal drift of the ANTARES microscope in ambient conditions. The left panel shows the time evolution   of the absolute position of the sample and the FZP, together with the temperature of the experimental hutch. Both interferometric signals confirm that the main perturbation is due to the thermal drift of the microscope. The right panel displays the position of the FZP in the X plane when it is scanned 140 $\mu$m along the beam incidence direction (s direction).}
\end{minipage}
\end{figure}

\section{First commissioning results}

The laser interferometer allows the direct measurement of the relative mechanical vibration and thermal drift of the FZP and the sample.  A representative set of measurements is displayed in Fig. 3. For experiments where the FZP nominal position is fixed (left panel of Fig. 3), spurious movements are of the order of micrometer or even larger. The time evolution of the interferometric signals indicates that the main instability is due to the microscope thermal drift. A cyclic temperature variation not greater than  0.15 $^{\circ }$C peak-to-peak produces a cyclic change of the absolute FZP and sample positions. This drift is similar in phase for both elements but different in amplitude and consequently should be compensated by the interferometric feedback. As the FZP is scanned along the incident beam direction, the positional variations are 50 $\%$ lower than the drift recorded at fixed mode FZP operation during a long period of time. However, when the interferometric system is operated in the closed-loop mode, vibrations and drifts in fixed and scanned operation modes can be reduced to about 5 nm peak-to-peak. The right side of Fig. 3, shows that the spurious movements of the FZP are precisely compensated by inducing the same positional displacement of the sample. Accordingly, even if the X-ray spot on the sample moves due to the change of the FZP position, the closed-loop piezo stage of the sample corrects the sample position in such a way that the portion of the illuminated specimen is always the same.  

These results suggest that the performance of the ANTARES microscope depends only  on the quality of the FZPs being utilized and the illumination conditions of the BL, in particular the  coherence of the X-ray beam to be focalized by the FZP.  

\section{Acknowledgments}
The authors acknowledge the support of the Engineering service of the Synchrotron SOLEIL.

\smallskip
 
\end{document}